\documentclass{ifacconf}

\usepackage{graphicx}      
\usepackage{natbib}        
\usepackage{amsmath,amssymb, bbm}
\usepackage{xcolor}
\usepackage{flushend}
\usepackage{bm}

\newcommand{\bs}{\boldsymbol}
\newcommand{\bx}{\bs x}

\newcommand{\bc}{\bs c}

\newcommand{\be}{\bs e}

\begin{document}
\begin{frontmatter}

\title{Synchronization of nonlinearly coupled networks of Chua oscillators\thanksref{footnoteinfo}} 

\thanks[footnoteinfo]{
This work has been supported by the Deutsche Forschungsgemeinschaft~(DFG) within the research unit FOR 2093: Memristive devices for neuronal systems (subproject C3: Synchronization of Memristively Coupled Oscillator Networks -- Theory and Emulation).
}

\author[First]{P. Feketa} 
\author[First]{A. Schaum} 
\author[First]{T. Meurer}
\author[Second]{D. Michaelis}
\author[Second]{K. Ochs}

\address[First]{Chair of Automatic Control, Christian-Albrechts-University Kiel, 24148 Kiel, Germany (e-mail: \{pf,alsc,tm\}@tf.uni-kiel.de).}
\address[Second]{Institute for Digital Communications Systems, Ruhr-University Bochum, 44801 Bochum, Germany (e-mail: \{dennis.michaelis, karlheinz.ochs\}@ruhr-uni-bochum.de)}

\begin{abstract} 
The paper develops new sufficient conditions for synchronization of a network of $N$ nonlinearly coupled Chua oscillators interconnected via the first state coordinate only. The nonlinear coupling strength is governed by a function residing within a sector, i.e. it is bounded from above and below by linear functions. The derived sufficient conditions provide a trade-off between the characteristics of the sector and the interconnection topology of the network to guarantee the synchronization of the  oscillators. 
\end{abstract}


\end{frontmatter}

\section{Introduction}
Control and synchronization of nonlinear and chaotic systems have been intensively studied during the last decades~\citep{pecora1997fundamentals, pecora2015synchronization, azar2017fractional, Ochs18, raychowdhury2019computing}. In particular, chaos synchronization has many potential applications in secure communication~\citep{yang1997impulsive, tse2003chaos, argyris2005chaos}, laser physics~\citep{ohtsubo2002chaos}, chemical reactor process~\citep{li2004experimental}, biomedical engineering~\citep{strogatz2018nonlinear}. In this context, the Chua circuit appeared to be one of the most interesting objects for research since it exhibits extremely rich dynamical behavior and variety of bifurcation phenomena despite its structural simplicity. Investigation of synchronization abilities of coupled Chua oscillators may help to understand complex dynamical phenomena arising in networks of chaotic systems of more general types~\citep{wu1994unified}.

Synchronization of two linearly coupled Chua oscillators has been studied in~\cite{wu1994unified, wang2006new, bowong2009practical, wang1999new, Zheng2002, chen2010chua, chen2012uncertain}. In~\cite{wu1994unified} it is conjectured that synchronization between two chaotic Chua circuits can be achieved by using the second state as the feedback variable for sufficiently large coupling constant. This conjecture has been analytically proven in~\cite{Zheng2002} and \cite{wang1999new} utilizing Lyapunov-type arguments and a novel observer design methodology.
Using the LaSalle invariance principle, a linear low-gain controller that exploits a single variable feedback (first state coordinate $x_1$) has been constructed in~\cite{chen2010chua}. The robustness of this controller with respect to the perturbations of the parameters of Chua  oscillators has been studied in~\cite{chen2012uncertain}. Practical synchronization of chaotic systems with uncertainties via adaptive coupling mechanism has been investigated in~\cite{bowong2009practical}. The synchronization of two identical chaotic and hyperchaotic systems with different initial conditions has been studied in~\cite{wang2006new}.
 
A graph-spectral approach for the synchronization of a network of $N\in\mathbb N$ resistively coupled nonlinear oscillators has been proposed in~\cite{WuChua1995}. The sufficient conditions for synchronization have been derived from the connectivity graph, which describes how the oscillators are connected. An upper bound on the coupling conductance required for synchronization for arbitrary graphs has been obtained.
Later, the synchronization of networks of $N\in\mathbb N$ nonlinear dynamical systems based on a state observer design approach has been studied in~\cite{Jiang2006}. Unlike the common diagonally coupling networks, see~\citep{wang2003complex, lu2004chaos}, where full state coupling is typically needed between two nodes, in~\cite{Jiang2006} it is suggested that only a scalar coupling signal is required to achieve network synchronization. The presented approach has been applied to the chaos synchronization problem in two typical dynamical network configurations: global linear coupling and nearest-neighbor linear coupling, with each node being a modified Chua's circuit. Finally, the problem of syhcnronizing an arbitrary subset of the nodes in the oscillatory network at fixed coupling stregth has been tackled in~\cite{gambuzza2018distributed} by creation of appropriate network of additional interconnecting links between oscillators.

In the current paper, the sufficient conditions for the synchronization of the network of $N\in\mathbb N$ Chua oscillators interconnected with the static nonlinear coupling via the first state coordinate only are derived. These conditions provide a trade-off between characteristics of the connectivity graph and properties of the nonlinear coupling function to achieve synchronization. 

The rest of the paper is organized as follows. In Section~2, the network under consideration is defined and the main problem of synchronization is formulated. In Section~3, the main result of the paper that is sufficient conditions for the synchronization of $N\in\mathbb N$ coupled Chua oscillators with static nonlinear coupling satisfying a so-called sector condition is derived. Also, numerical examples to illustrate the usage of the derived conditions are provided. In Section~4, a corollary of the main theorem for the case of two linearly coupled Chua oscillators is discussed and compared with the existing results in the literature. Finally, short conclusion and discussion in Section~5 complete the paper.

\section{Problem Statement}

Consider the extended Chua circuit system
\begin{subequations}\label{Chua}
	\begin{align}
	\dot x_1 &= \alpha(-x_1+x_2-f(x_1)){+ u}, & x_1(0)=x_{10}\\
	\dot x_2 &= x_1-x_2+x_3, & x_2(0)=x_{20}\\
	\dot x_3 &= -\beta x_2-\gamma x_3, & x_3(0)=x_{30}\\
	y &= x_1
	\end{align}
\end{subequations}
with scalar piecewise linear function $$f({x}) = ax+\frac{1}{2}(b-a)(|x+1|-|x-1|)$$ and parameters $\alpha,\beta>0,\gamma\geq0$, $a<b<0$. The system allows for vector-valued formulation
\begin{subequations}\label{Sys}
	\begin{align}
	\dot \bx &= \bs{A}\bx + {\bs{b}u} + \bs{f}(\bs{x}), & \bx(0)=\bx_0\\
	y &= \bc^T\bx
	\end{align}
\end{subequations}
with the state $\bx{(t)}\in\mathbb R^3$, {$\bs{x}_0^T=\begin{bmatrix}x_{10} & x_{20} & x_{30}\end{bmatrix}$}{, external input $u(t)\in\mathbb R$, which will be later used to interconnect Chua oscillators,} and the matrix $\bs{A}$ and vectors $\bs{f},\bs{b},\bc$ given by
\begin{align*}
\bs{A}=\begin{bmatrix}
-\alpha & \alpha & 0 \\ 1 & -1 & 1 \\ 0 & -\beta & -\gamma 
\end{bmatrix},
\,
\bs{f}(\bs{x}) = \begin{bmatrix} -\alpha f(x_1) \\ 0 \\ 0 \end{bmatrix},
\,
\bs{b} = \bc = \begin{bmatrix} 1 \\ 0 \\ 0\end{bmatrix}.
\end{align*}

Now, consider a network of $N\in\mathbb N$ nodes described by a graph $\Gamma = (V,E)$ with vertex (node) set $V$ and edge set $E$, so that $|V|=N$. Let the associated adjacency matrix be given by $\bm{\mathcal A}=\{\alpha_{ij}\}_{i,j=1,\ldots,N}$, $\alpha_{ij}\in\left\{0,1 \right\}$ with zero main diagonal. 

Then, applying output feedback
\begin{align*}
	u_i= -\sum_{j=1}^{N} \alpha_{ij}k(y_i-y_j), \quad i=1,\ldots,N
\end{align*}
with an arbitrary nonlinear locally Lipschitz continuous coupling function $k:\mathbb R\to\mathbb R$, the dynamics of $N\in\mathbb N$ coupled Chua oscillators can be written as 
\begin{equation}\label{N-Chua}
	\begin{aligned}
	\dot\bx_i &= \bs{A}\bx_i -\bs{b}\sum_{j=1}^N \alpha_{ij}k(y_i-y_j) +{\bs{f}(\bs{x_i})},\quad \bx_i(0)=\bx_{i 0}\\
	y_i &= \bc^T\bx_i, 
	\end{aligned}
\end{equation}
for ${i}=1,\ldots,N$. 
For any given $\bs\xi\in\mathbb R^{3N}$ let $\bx=(\bx_1,\ldots,\bx_N): \mathbb R \to \mathbb R^{3N}$ denote a solution to \eqref{N-Chua} satisfying the initial condition $\bx(0)=\bs\xi$. The Lipschitz continuity of the right-hand side of (3) guarantees the existence and uniqueness of the solution for any initial value $\bs\xi\in\mathbb R^{3N}$.

Associated to this network consider the relative synchronization errors with respect to the node 1
\begin{align*}
\be_j = \bx_j-\bx_1,\quad {j}=1,\ldots,N.
\end{align*}
The relative errors $\be_{ij}$ between arbitrary nodes $i$ and $j$ can be expressed using the relative errors $\be_j$ and $\be_i$
\begin{align*}
\be_{ij} =  \bx_i-\bx_j = \bx_i-\bx_1 - (\bx_j-\bx_1) = \be_i-\be_j.
\end{align*}
Accordingly, instead of analyzing $\frac{N(N-1)}{2}$ relative errors $\be_{ij}$ between connected nodes, it is sufficient to consider the behavior of the $N{-1}$ errors $\be_j,\, {j=2},\ldots,N$.

The problem addressed in the sequel consists in providing sufficient conditions on the system parameters, {the nonlinear coupling function $k$ and the network topology} which ensure the synchronization of $N\in\mathbb N$ coupled Chua oscillators, i.e., the global convergence of the norms of errors $\be_j$ to zero:
\begin{align*}
\lim_{t\to\infty}\|\be_j(t)\|=0,\quad {j=2},\ldots,N.
\end{align*}

\section{Synchronization Conditions}



Let $\{\bx\}_i$ denote the $i$-th component of the vector $\bx\in\mathbb R^3$, $i=1,2,3$.
By splitting the state vector according to 
\begin{align*}
 \begin{bmatrix} z_i\\\bs\zeta_i\end{bmatrix} = \begin{bmatrix} \{\bx_i\}_1 \\ \begin{bmatrix}\{\bx_i\}_2\\ \{\bx_i\}_3\end{bmatrix}\end{bmatrix},
\end{align*}
rewrite the dynamics \eqref{N-Chua} in ${z}-\bs\zeta$ coordinates
\begin{subequations}\label{N-Chua2}
\begin{align}
 \dot z_i &= -\alpha z_i - \alpha f(z_i) + \begin{bmatrix} \alpha & 0\end{bmatrix} \bs\zeta_i \\ \nonumber
 & \quad- \sum_{j=1}^N \alpha_{ij} k(z_i-{z_j})
 \\
 \dot{\bs{\zeta}}_i &= \begin{bmatrix} 1\\0\end{bmatrix}z_i + \underbrace{\begin{bmatrix}
 -1 & 1\\-\beta & -\gamma
 \end{bmatrix}}_{=:\bs{A}_0} \bs\zeta_i,
\end{align}
\end{subequations}
where matrix ${\bs{A}_0}$ is Hurwitz with eigenvalues $\lambda_{1,2}$ fulfilling $\mathfrak R(\lambda_{1,2}) = -\mu_0<0$ with
\begin{align*}
{-\mu_0=-\frac{1+\gamma}{2} + \mathfrak R\left(\sqrt{\frac{(1+\gamma)^2}{4}-(\gamma+\beta)}\right)},
\end{align*}
{where $\mathfrak R(\lambda)$ denotes the real part of $\lambda\in\mathbb C$.} 


Using the notation for the relative errors
\begin{align*}
 \begin{bmatrix} e_i\\\bs\eta_i\end{bmatrix} = \begin{bmatrix} z_i-z_1 \\ \bs\zeta_i-\bs\zeta_1\end{bmatrix}
\end{align*}
{and taking into account that $e_1\equiv 0$, $\bs\eta_1 \equiv \bs 0$, the synchronization error dynamics can be written as}
\begin{subequations}\label{Error}
\begin{align}\label{Error1}
 \dot e_i &= -\alpha e_i - \alpha {\tilde{f}(e_i)}+ \begin{bmatrix} \alpha & 0\end{bmatrix} \bs\eta_i \\ \notag & \quad- \sum_{j=1}^N \Big(\alpha_{ij} k(e_i-e_j) - \alpha_{1j}k(-e_j)\Big) \\
 \label{Error2}
 \dot{\bs{\eta}}_i &= \begin{bmatrix} 1\\0\end{bmatrix}e_{i} + A_0\bs\eta_i
\end{align}
\end{subequations}
with initial conditions $e_i(0)=e_{i,0}$ and $\bs{\eta}_i(0)=\bs{\eta}_{i,0}$, $i=2,\ldots,N$, and
\begin{equation}\label{f_tilde}
\tilde{f}(e_i) = f(z_1+e_i)-f(z_1).
\end{equation} 

In the following subsection, sufficient conditions for global asymptotic stability of zero solution to the error dynamics system \eqref{Error} will be derived. For this purpose additional sector requirement on the coupling function $k$ will be imposed.

\subsection{Coupling with sector condition}

\begin{assum}\label{assumpt1}
Coupling $k:\mathbb R \to \mathbb R$ is continuous odd function and there exist two constants $k_2 \geq k_1 \geq 0$ such that for all $e\in\mathbb R$:
\begin{equation}\label{sector_cond}
k(e)e\geq 0, \quad k(-e)=-k(e), \quad k_1|e|\leq |k(e)|\leq k_2|e|.
\end{equation}
\end{assum}


Introduce the function
\begin{align}\label{ktilde}
 \tilde k(e) = k(e)-k_2e \quad \text{for all}\quad e\in\mathbb R.
\end{align}
From \eqref{sector_cond} and \eqref{ktilde} it follows that functions $k$ and $\tilde k$ lie in the first-third and the second-fourth quadrant pairs respectively.


\begin{lem}\label{lemmaK}
Let Assumption~\ref{assumpt1} hold. Then
\begin{equation}\label{eqlemma}
\left |\tilde k(e_i-e_j)+\tilde k(e_j) \right | \leq \left(k_2-k_1 \right)|e_i|+\left(k_2-k_1\right)|e_j|
\end{equation}
for all $e_i, e_j \in \mathbb R$.
\end{lem}
\begin{pf}
The left-hand side of \eqref{eqlemma} can be rewritten as
\begin{equation*}
\begin{split}
\left |\tilde k(e_i-e_j)+\tilde k(e_j) \right | & = |k(e_i-e_j) - k_2(e_i-e_j) \\& \quad+ k(e_j) - k_2 e_j | \\
& = \left | -k_2e_i + k(e_j)-k(e_j-e_i)\right |
\end{split}
\end{equation*}
Consider the cases of positive and negative signs of the terms $e_i, e_j$, and $e_j-e_i$ correspondingly. First, let $e_i\geq 0$, $e_j\geq 0$, $e_j-e_i\geq 0$. Then,
\begin{align*}
-k_2e_i+k_1e_j-k_2(e_j-e_i) &\leq  -k_2e_i + k(e_j)-k(e_j-e_i) \\&\leq -k_2e_i+k_2e_j-k_1(e_j-e_i),
\end{align*}
which yields
\begin{equation}\label{est1}
\begin{aligned}
|-k_2e_i + k(e_j)-k(e_j-e_i)| \leq (k_2-k_1)|e_j|.
\end{aligned}
\end{equation}
Let $e_i\geq 0$, $e_j\geq 0$, $e_j-e_i< 0$. Then,
\begin{align*}
-k_2e_i+k_1e_j-k_1(e_j-e_i) &\leq  -k_2e_i + k(e_j)-k(e_j-e_i) \\&\leq -k_2e_i+k_2e_j-k_2(e_j-e_i) \\ &\leq 0,
\end{align*}
which yields
\begin{align}\label{est2}
|-k_2e_i + k(e_j)-k(e_j-e_i)| \leq (k_2-k_1)|e_i|.
\end{align}
Let $e_i\geq 0$, $e_j< 0$, $e_j-e_i< 0$. Then,
\begin{align*}
-k_2e_i+k_2e_j-k_1(e_j-e_i) &\leq  -k_2e_i + k(e_j)-k(e_j-e_i) \\&\leq -k_2e_i+k_1e_j-k_2(e_j-e_i)
\end{align*}
which yields
\begin{equation}\label{est3}
\begin{aligned}
|-k_2e_i + k(e_j)&-k(e_j-e_i)|  \\ &\leq (k_2-k_1)|e_i|+(k_2-k_1)|e_j|.
\end{aligned}
\end{equation}
Similarly, one may check that for the rest combinations of the signs of $e_i$, $e_j$, and $e_j-e_i$ one of the inequalities \eqref{est1}, \eqref{est2}, \eqref{est3} holds. Finally, combining \eqref{est1}, \eqref{est2}, \eqref{est3}, obtain that
\begin{equation*}
\left |\tilde k(e_i-e_j)+\tilde k(e_j) \right | \leq \left(k_2-k_1 \right)|e_i|+\left(k_2-k_1\right)|e_j|.
\end{equation*}
This completes the proof.\hfill$\square$
\end{pf}

Denote the degree of node $i$ by $\kappa_i=\sum\limits_{j=1}^N \alpha_{ij}$ and rewrite the dynamics \eqref{Error1} as follows:
\begin{equation}\label{tempsum}
\begin{aligned}
 \dot e_i =& -\alpha e_i - \alpha \tilde f(e_i) + \begin{bmatrix} \alpha & 0\end{bmatrix} \bs\eta_i \\ &- \sum_{j=1}^N \Big(\alpha_{ij} k(e_i-e_j) + \alpha_{1j}k(e_j)\Big)\\
  =& - \alpha e_i-\kappa_i k_2 e_i - \alpha \tilde f(e_i) + \begin{bmatrix} \alpha & 0\end{bmatrix} \bs\eta_i \\
  &- \sum_{j=1}^N \Big(\alpha_{ij} \big(k(e_i-e_j)-k_2 (e_i-e_j)\big) \\& + \alpha_{1j}k(e_j)-\alpha_{ij}k_2 e_j \Big).  
\end{aligned}
\end{equation}

Introduce the residual connectivity coefficients with respect to the first node 
\begin{align*}
 \alpha_{1j}&=\alpha_{ij}+\tilde \alpha_{ij},\\ \tilde \alpha_{ij} &= \alpha_{1j}-\alpha_{ij}{=
 \begin{cases}
    0,&(i,j),(1,j)\in E,\\
    1,&(i,j)\notin E,(1,j)\in E,\\
    -1,&(i,j)\in E,(1,j)\notin E, \\
    0,&(i,j),(1,j)\notin E,
\end{cases}}
\end{align*}
respectively. The expression inside the sum of~\eqref{tempsum} can be rewritten as
\begin{align*}
 \alpha_{ij}\Big(k(e_i&-e_j)-k_2(e_i-e_j)\Big)+\alpha_{1j}k(e_j)-\alpha_{ij}k_2 e_j\\ 
 &= \alpha_{ij}\Big(k(e_i-e_j)-k_2(e_i-e_j)\Big) \\ &\quad+(\alpha_{ij}+\tilde\alpha_{ij}) k(e_j) -\alpha_{ij}k_2 e_j \\   
 &= \alpha_{ij}\Big(\tilde k(e_i-e_j)+\tilde k(e_j)\Big)+\tilde \alpha_{ij} k(e_j). 
\end{align*}
With these definitions the dynamics of $e_i$ can be written equivalently as
\begin{equation}\label{precodes}
\begin{aligned}
 \dot e_i =& -\alpha e_i-\kappa_i k_2 e_i - \alpha \tilde f(e_i) + \begin{bmatrix} \alpha & 0\end{bmatrix} \bs\eta_i \\ &- \sum_{j=1}^N \Big(\alpha_{ij} \big(\tilde k(e_i-e_j)+\tilde k(e_j)\big) + \tilde\alpha_{ij} k(e_j)\Big).
\end{aligned}
\end{equation}
{Following the reasoning in \citep{SCHAUM201846} consider the implicit solution of the preceding ODEs \eqref{Error2}, \eqref{precodes} given by}
\begin{equation}\label{newEq}
\begin{aligned}
 e_i(t) &= e^{-(\alpha+\kappa_i k_2)t}e_{i,0} \\ +&\int_0^t e^{-(\alpha+\kappa_i k_2)(t-\tau)} \bigg(-\alpha\tilde f(e_i(\tau))+\begin{bmatrix} \alpha & 0\end{bmatrix} \bs\eta_i(\tau)\\
          {-}&\sum_{j=1}^N \left(\alpha_{ij} \left(\tilde k(e_i-e_j)+\tilde k(e_j)\right) + \tilde\alpha_{ij}k(e_j)\right)\bigg)d\tau\\
 \bs\eta_i(t) &= e^{\bs{A}_0 t}\bs \eta_{i,0} + \int_0^t e^{\bs{A}_0 (t-\tau)}\begin{bmatrix} 1\\0\end{bmatrix}e_i(\tau)d\tau.
\end{aligned}
\end{equation}

Since function $f$ is Lipschitz continuous with Lipschitz constant $|a|$ it holds that
\begin{equation}\label{eqtildef}
|\tilde f(e_i)| = |f(z_1+e_i)-f(z_1)| \leq |a||e_i|
\end{equation}
for any $e_i\in\mathbb R$. Taking norms on both sides of \eqref{newEq}, applying the triangle inequality, and accounting for Lemma~\ref{lemmaK} and inequality \eqref{eqtildef}
the estimates
\begin{align*}
 |e_i(t)| &\leq e^{-(\alpha+\kappa_i k_2)t} \\ &\times \left(|e_{i,0}| + \int_0^t {e^{(\alpha+\kappa_i k_2)\tau}} \Bigl( \alpha |a||e_i(\tau)|+\alpha\|\bs\eta_i(\tau)\| \right.\\
    &
     \left.{+}\sum_{j=1}^N \left(\alpha_{ij} (\left(k_2-k_1 \right)|e_i(\tau)|+\left(k_2-k_1\right)|e_j(\tau)|) \right.\right. \\& \left.\left.{+} |\tilde\alpha_{ij}|k_2|e_j(\tau)|\right) \Bigr) d\tau\right)\\
 \|\bs\eta_i(t)\| &\leq e^{-\mu_0 t}\left(\|\bs\eta_{i,0}\| + \int_0^t {e^{\mu_0 \tau}}|e_i(\tau)|d\tau\right)
\end{align*}
hold. {Define the right-hand sides of the preceding inequalities as $\sigma_i$ and $\chi_i$, i.e.,
\begin{align*}
 \begin{split}
 \sigma_i &= e^{-(\alpha+\kappa_i k_2)t} \\ &\times\left(|e_{i,0}| + \int_0^t {e^{(\alpha+\kappa_i k_2)\tau}} \Big( \alpha |a||e_i(\tau)|+\alpha\|\bs\eta_i(\tau)\| \right.\\
    &
      \left.{+}\sum_{j=1}^N \left(\alpha_{ij} (\left(k_2-k_1 \right)|e_i(\tau)|+\left(k_2-k_1\right)|e_j(\tau)|) \right.\right. \\&\left.\left.+ |\tilde\alpha_{ij}|k_2|e_j(\tau)|\right) \Big) d\tau\right)
 \end{split} 
 \\
 \chi_i &= e^{-\mu_0 t}\left(\|\bs\eta_{i,0}\| + \int_0^t {e^{\mu_0 \tau}}|e_i(\tau)|d\tau\right)   
\end{align*}
so that $|e_i(t)|\leq\sigma_i(t)$ and $\|\bs\eta_i(t)\|\leq\chi_i(t)$ for all $t\geq 0$ with $|e_i(0)|=\sigma_i(0)$ and $\|\bs\eta_i(0)\|=\chi_i(0)$ for all $i=2,\ldots,N$. Since $e_1(t)=0$ for all $t\geq 0$, let $\sigma_1(t)\equiv 0$ so that $|e_i(t)|\leq\sigma_i(t)$ holds for all $i=1,\ldots,N$. The time derivatives of $\sigma_i$ and $\chi_i$ can be estimated by

\begin{align*}
 \begin{split}
 \dot\sigma_i(t) =& -(\alpha+\kappa_i k_2)\sigma_i(t) + \alpha |a| |e_i(t)|+\alpha \|\bs\eta_i(t)\|
    \\ & + \sum_{j=1}^N \big(\alpha_{ij}
	\left(    
    \left(k_2-k_1 \right)|e_i|+\left(k_2-k_1\right)|e_j|
	\right)    
    \\ &+ |\tilde\alpha_{ij}|k_2|e_j(t)|\big) \\
    \leq&
    -(\alpha-\alpha |a|+\kappa_i k_1)\sigma_i(t) + \alpha \chi_i(t) \\&+\sum_{j=1}^N \left( |\alpha_{ij}|(k_2-k_1) + |\tilde\alpha_{ij}|k_2 \right)\sigma_j(t)    
\end{split}  
\\
\begin{split} 
 \dot \chi_i(t) =&-\mu_0\chi_i(t)+|e_i(t)| \\ \leq& -\mu_0\chi_i(t) + \sigma_i(t).
 \end{split}
\end{align*} 
{Taking into account that $\sigma_1(t)\equiv 0$,} the preceding \mbox{dynamics} can be written in vector notation as 
\begin{equation}\label{tempeq}  
\begin{aligned}
 \frac{d}{dt} \begin{bmatrix} \sigma_i\\\chi_i\end{bmatrix} 
 &\leq 
 \begin{bmatrix} -(\alpha-\alpha|a|+\kappa_i k_1) & \alpha \\ 1 & -\mu_0\end{bmatrix}\begin{bmatrix} \sigma_i\\\chi_i\end{bmatrix} 
\\ &+ \begin{bmatrix}\sum_{j=2}^N \left( |\alpha_{ij}|(k_2-k_1)+|\tilde \alpha_{ij}|k_2 \right)\sigma_j\\ 0\end{bmatrix}.
\end{aligned}
\end{equation}
Introducing $\bs z = \begin{bmatrix} \sigma_2 & \cdots & \sigma_N& \chi_2 & \cdots & \chi_N\end{bmatrix}^T$, inequality \eqref{tempeq} can be written as
\begin{align}\label{def_M}  
 \dot{\bs{z}} &\leq \underbrace{\begin{bmatrix} -(\alpha-\alpha|a|) \mathbf{I} -k_1 \mathbf{K} + (k_2-k_1)\bm{\mathcal{A}}_1 + k_2 \bm{\mathcal{A}}_2 & \alpha \mathbf{I} \\ \mathbf{I} & -\mu_0 \mathbf{I} \end{bmatrix}}_{=:\mathbf M}\bs z 
\end{align}  
with matrices
\begin{align*}
  \bm{\mathcal A}_1 &= 
    \begin{bmatrix}
     0 & |\alpha_{23} | & \cdots &|\alpha_{2N}|\\
     |\alpha_{32} | & 0 & \ddots & \vdots\\
     \vdots & \ddots & \ddots & \vdots \\
     |\alpha_{N2} | & \cdots & \cdots & 0
    \end{bmatrix},
    \\
  \bm{\mathcal A}_2 &= 
    \begin{bmatrix}
     0 & |\alpha_{23} -\alpha_{13}| & \cdots &|\alpha_{2N}- \alpha_{1N}|\\
     |\alpha_{32} - \alpha_{12}| & 0 & \ddots & \vdots\\
     \vdots & \ddots & \ddots & \vdots \\
     |\alpha_{N2} -\alpha_{12}| & \cdots & \cdots & 0
    \end{bmatrix}, 
\end{align*}           
$\mathbf K=\text{diag}\{\kappa_2,\ldots,\kappa_N\}$, and $(N-1)\times(N-1)$--dimensional identity matrix $\mathbf I$. Sufficient conditions for the synchronization of the entire network can be formulated in terms of the eigenvalues of the matrix $\mathbf M$.

\begin{thm}\label{Thm:Main2}
Let Assumtion~\ref{assumpt1} hold and matrix $\mathbf M$ defined in \eqref{def_M} be Hurwitz. Then, the norm of errors between the states of Chua oscillators \eqref{N-Chua} converges exponentially to zero.
\end{thm}

The differential inequality \eqref{tempeq} for $\sigma_i$ contains both stabilizing and destabilizing terms, which have physical interpretation and can be used as guidelines for the coupling design. In particular, the stabilizing term becomes larger with the growth of the lower bound $k_1$ of the nonlinear coupling. The destabilizing terms vanish when the lower bound $k_1$ approaches the upper bound $k_2$ and the interconnection graph is fully connected (i.e. the coefficients $\tilde \alpha_{ij}$ are zero). These effects can be reached by choosing the coupling $k$ with a sufficiently large lower bound $k_1$.


\subsection{Numerical example}

\subsubsection{Example 1.}
Consider a fully connected network of $N=20$ Chua oscillators \eqref{N-Chua} with parameters $\alpha = 15.61$, $\beta=25.581$, $\gamma=0$, $a=-1.142$, $b=-0.715$,
and nonlinear coupling
\begin{equation}\label{kkk}
k(e)=3e+\arctan{(e)}\quad \text{for all}\quad e\in\mathbb R.
\end{equation}
The chosen parameters correspond to the chaotic behavior of each oscillator~\citep{Pivka1994}. Oscillators' trajectories converge to an attractor that has a double scroll shape in three dimensional state space (see Fig.~\ref{Fig:doubleScroll}).
\begin{figure}[!h]
 \centering
  \includegraphics[width=0.5\textwidth]{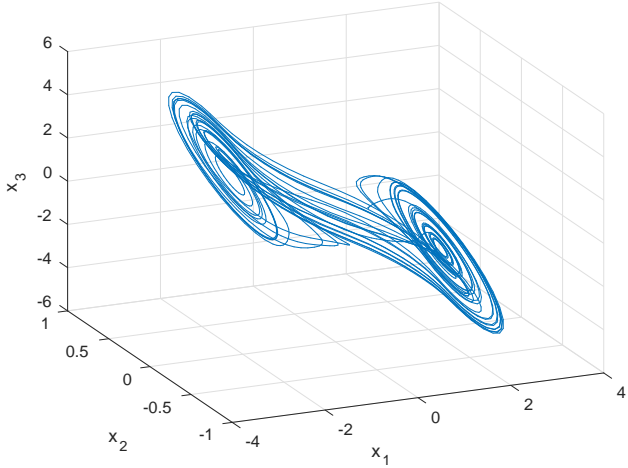}
 \caption{Double scroll attractor for the coupled chaotic Chua oscillators from Example~1.}
 \label{Fig:doubleScroll}
\end{figure}

Nonlinear coupling strength \eqref{kkk} satisfies the sector condition \eqref{sector_cond} with constants $k_1=3$ and $k_2=4$. For the chosen parameters and the interconnection coupling \eqref{kkk} the matrix $\mathbf M$ defined in \eqref{def_M} reads

\begin{equation*}
\mathbf M=\begin{bmatrix}
-54.7834 \mathbf I
&
15.61 \mathbf I
\\
\mathbf I
&
-0.5 \mathbf I
\end{bmatrix}
+
\begin{bmatrix}
\mathbf{1} {\mathbf{1}}^\top - \mathbf I
&
\mathbf{0}
\\
\mathbf{0}
&
\mathbf{0}
\end{bmatrix},
\end{equation*}
where $\mathbf{0}$ denotes zero $(N-1)\times(N-1)$--matrix, and $\mathbf{1}$ denotes $(N-1)$--dimensional vector $\begin{bmatrix}1 & 1 & \cdots & 1 \end{bmatrix}^\top$. By direct calculation one may check that all eigenvalues lie in the open left half-plane
\begin{align*}
\mathfrak R(\lambda(\mathbf M)) \in [-56.0643, -0.0748].
\end{align*}
Hence, from Theorem~\ref{Thm:Main2} it follows that the oscillators are synchronized. The state evolution for $N=20$ oscillators is shown in Fig.~\ref{Fig:states20Osc}.
\begin{figure}[!h]
 \centering
      \includegraphics[width=0.5\textwidth]{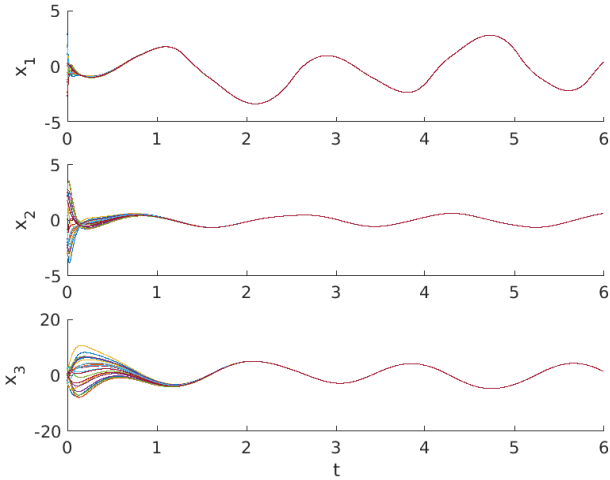}
 \caption{Time evolution of $N=20$ coupled chaotic Chua oscillators with interconnection coupling \eqref{kkk}.}
 \label{Fig:states20Osc}
\end{figure}

\pagebreak
\section{Synchronization of two oscillators}

In this section a corollary from Theorem~\ref{Thm:Main2} for the case of two Chua oscillators connected with linear coupling 
\begin{equation}\label{lincoupl}
k(e)=ke \quad \text{for all} \quad e\in \mathbb R
\end{equation}
with coupling constant $k>0$ is presented. Obviously, the function \eqref{lincoupl} satisfies the sector condition \eqref{sector_cond} with $k_1=k_2=k$.



\begin{cor}\label{cor1}
In the case of two Chua oscillators \eqref{N-Chua} connected via the first state variable with linear coupling \eqref{lincoupl} the condition for synchronization reads
\begin{equation}\label{twolin}
k>\alpha \left( |a| + \frac{1}{\mu_0} -1 \right),
\end{equation}
i.e., the synchronization emerges if the coupling between the first state of each oscillator is sufficiently strong.
\end{cor}
\begin{pf}
For the case of two Chua oscillators the matrix $\mathbf M$ from \eqref{def_M} is a $2\times2$--matrix defined by
\begin{align*}
\mathbf M=\begin{bmatrix} -(\alpha-\alpha|a|+ k) & \alpha \\ 1 & -\mu_0 \end{bmatrix}.
\end{align*}  
Due to the Routh-Hurwitz criterion, the roots of the corresponding characteristic polynomial
\begin{align*}
\lambda^2+\lambda\left(\mu_0+\alpha-\alpha|a|+k\right)+\left(\alpha-\alpha|a|+k \right)\mu_0-\alpha=0
\end{align*}  
are in the open left half-plane if and only if
\begin{align*}
\begin{cases}
\mu_0+\alpha-\alpha|a|+k > 0, \\
\left(\alpha-\alpha|a|+k \right)\mu_0-\alpha > 0,
\end{cases}
\end{align*}  
which yields 
\begin{equation*}
k>\alpha \left( |a| + \frac{1}{\mu_0} -1 \right).
\end{equation*}
This completes the proof.\hfill$\square$
\end{pf}

The problem of the synchronization of two identical Chua oscillators via the first state variable by linear coupling addressed in Corollary~\ref{cor1} has been also successfully tackled in~\cite{chen2010chua, chen2012uncertain}  for the case of $\gamma=0$ by employing the Lyapunov method. The constraint on the coupling constant obtained in~\cite{chen2010chua, chen2012uncertain} reads as
\begin{equation*}
k>\alpha |a|,
\end{equation*}
which is less a conservative condition compared to \eqref{twolin}. However, the applicability of Theorem~\ref{Thm:Main2} is more general even in the case of two oscillators due to the possibility of $\gamma\not = 0$. Besides this, the approach proposed in the present paper handles the case of an arbitrary number of Chua oscillators with an arbitrary interconnection topology and nonlinear coupling functions.

\subsubsection{Example 2.}

Consider two Chua oscillators with parameters $\alpha = 10$, $\beta=15$, $\gamma=0.1$, $a=-1.31$, $b=-0.75$,
which are linearly coupled via the first state coordinate. By direct calculation we obtain that
\begin{equation*}
-\mu_0=-\frac{1.1}{2}+\mathfrak R\left(\sqrt{\frac{1.21}{4}-15}\right)=-\frac{1.1}{2}.
\end{equation*}
From Corollary~\ref{cor1}, the coupling constant $k$ should be chosen larger than
\begin{equation*}
k>10\left( 1.31+\frac{2}{1.1}-1\right)\approx 21.282.
\end{equation*}

The state evolution of both oscillators for $k=21.3$ is shown in Fig.~\ref{Fig:states2Osc}.
\begin{figure}[!h]
 \centering
   \includegraphics[width=0.5\textwidth]{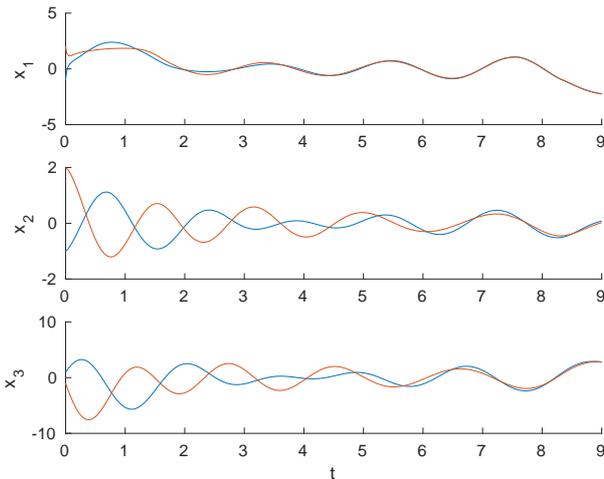}
 \caption{Time evolution of two coupled chaotic Chua oscillators with interconnection gain $k=21.3$.}
 \label{Fig:states2Osc}
\end{figure}

\section{Conclusion and Outlook}

The synchronization of a network of $N\in\mathbb N$ Chua oscillators which are coupled via the first state coordinate with static nonlinear coupling is studied. Sufficient conditions for the synchronization are formulated in terms of the eigenvalues of the auxiliary matrix $\mathbf M$, whose entries represent the interplay between the parameters of the oscillators, the interconnection topology of the network and the characteristics of the nonlinear coupling function. 

The extension of the derived conditions to the class of dynamically coupled Chua oscillators will allow for analysis of wide classes of memristive networks and, more generally, time-varying interconnections which are capable to model the bio-inspired plasticity phenomenon. Another interesting research direction is the study of multi-clustering capabilities of oscillatory networks and control design approaches for this kind of behaviour.





\bibliography{ifacconf}             
                                                   







\end{document}